**Open-Source Tools for Behavioral Video Analysis: Setup, Methods, and Best Practices**

Kevin Luxem[1]*, Jennifer J. Sun[2]*, Sean P. Bradley[3], Keerthi Krishnan[4], Eric A. Yttri[5], Jan Zimmermann[6], Talmo D. Pereira[7], and Mark Laubach[8]

[1]Cellular Neuroscience, Leibniz Institute for Neurobiology, Magdeburg, Germany
[2]Department of Computing & Mathematical Sciences, California Institute of Technology, California, USA
[3]Rodent Behavioral Core, National Institute of Mental Health, National Institutes of Health, Bethesda, MD, USA
[4]Department of Biochemistry and Cellular & Molecular Biology, University of Tennessee, Knoxville, TN, USA
[5]Department of Biological Sciences, Carnegie Mellon University, Pittsburgh, PA, USA
[6]Department of Neuroscience, University of Minnesota, Minneapolis, MN, USA
[7]The Salk Institute of Biological Studies, La Jolla, CA 92037, USA
[8]Department of Neuroscience, American University, Washington, DC, USA

* = Co-first author

Corresponding authors:
Mark Laubach, PhD
Department of Neuroscience
American University
4400 Massachusetts Ave NW
Washington, DC 20016
mark.laubach@american.edu

Talmo Pereira, PhD
The Salk Institute of Biological Studies
10010 N Torrey Pines Rd
La Jolla, CA 92037
talmo@salk.edu

Acknowledgments: This paper emerged from a working group on methods for video analysis organized by the OpenBehavior project in the summer and fall of 2021. Ann Kennedy, Greg Corder, and Sam Golden were major contributors to the working group and their ideas impacted this manuscript. We would like to thank Ann Kennedy, Samantha White and Jensen Palmer for helpful comments on the manuscript.

Conflict of Interest: None

Financial Support: NSERC Award #PGSD3-532647-2019 to JJS; NIH MH002952 for SPB; NIH MH124042 for KK; NIH MH128177 and NSF 2024581 to JZ; NSF 1948181 and NIH DA046375 to ML

**Abstract**

Recently developed methods for video analysis, especially models for pose estimation and behavior classification, are transforming behavioral quantification to be more precise, scalable, and reproducible in fields such as neuroscience and ethology. These tools overcome long-standing limitations of manual scoring of video frames and traditional "center of mass" tracking algorithms to enable video analysis at scale. The expansion of open-source tools for video acquisition and analysis has led to new experimental approaches to understand behavior. Here, we review currently available open-source tools for video analysis and discuss how to set up these methods for labs new to video recording. We also discuss best practices for developing and using video analysis methods, including community-wide standards and critical needs for the open sharing of datasets and code, more widespread comparisons of video analysis methods, and better documentation for these methods especially for new users. We encourage broader adoption and continued development of these tools, which have tremendous potential for accelerating scientific progress in understanding the brain and behavior.

**Quantitative tools for video analysis**

Traditional approaches to analyzing video data have involved researchers watching video playback and noting the times and locations of specific events of interest. These analyses are very time-consuming, require expert knowledge in the target species and experimental design, and are prone to user bias (Anderson and Perona, 2014). Video recordings are often made for many different animals and behavioral test sessions, but only reviewed for a subset of experiments. Complete sets of videos are rarely made accessible in published studies and the analysis methods are often vaguely described. There are variations in scoring criteria across researchers and labs, even over time by a single researcher. Collectively, these issues present major challenges for research reproducibility and the difficulty and cost of manual video analysis has led to the dominance of easy-to-use measures (lever pressing, beam breaks) in the neuroscience literature, and this has limited our understanding of brain-behavior relationships (Krakauer et al., 2017).

For example, "reward seeking" has been a popular topic in recent years and is typically measured using beam breaks between response and reward ports located inside an operant arena (e.g., Cowen et al., 2012; Feierstein et al., 2006; Lardeux et al., 2009; van Duuren et al., 2009). By relying only on the discrete times when animals make a choice and receive a reward, it is not possible to describe how the animal moves during a choice or how it collects a reward. Animals may not move in the same way to a reward port when they expect a larger or smaller reward (e.g., Davidson et al., 1980). This could lead to, for example, a neural recording study labeling a cell as "reward encoding" when it actually reflects differences in movement.

Commercial products (e.g., Ethovision by Noldus, Any-Maze by Stoelting) and open-source projects (e.g., JAABA: Kabra et al., 2013; SCORHE: Salem et al., 2015; OptiMouse: Ben-Shaul, 2017; ezTrack: Pennington et al., 2019) are available for semi-automated annotation and tracking of behaviors. These methods track animals based on differences between the animals and the background color or luminance. This can be challenging to do in naturalistic settings or for species or strains that do not have a uniform color (e.g., Long-Evans rats). These methods provide estimates of the overall position of an animal in its environment and can be used to measure the direction and velocity of its movements. These "center of mass" tracking methods could be used to measure where an animal is and how fast it is moving. More sophisticated versions of these products may also detect the head and tail of common laboratory species such as rodents or zebrafish and draw inferences from the shape and location of the animal to classify a small subset of an animal's behavioral repertoire. However, these simpler tracking methods cannot account for movements of discrete sets of body parts (e.g., head scanning in rodents, which is associated with a classic measure of reward-guided decisions called "vicarious trial-and-error" behavior: see Redish, 2016 for review).

More advanced analyses could be used to quantify movements across many pixels simultaneously in video recordings. For example, Stringer et al. (2019) used dimensionality reduction methods to study the spontaneous coding of visual and movement related information in the mouse visual cortex in relation to facial movements. Musall et al. (2019) used video recordings of motion data from several parts of the face of mice as they performed a decision-making task and related the measures from the video recordings to cortical imaging data. While these analyses would go beyond what is possible to achieve with a simple tracking method, the multivariate methods developed by Stringer and Musall are not themselves capable of categorizing movements, measuring transitions between different types of movements, or quantifying the dynamics of movement sequences. For these measures, a different approach is needed.

Methods for capturing the pose of an animal (the location and configuration of its body) have emerged in recent years (e.g., DeepLabCut: Mathis et al., 2018; SLEAP: Pereira et al., 2022). These methods can provide a description of an animal's movement and posture during a behavioral task. They can be used to understand the dynamics of naturalistic movements and behaviors, as illustrated in Figure 1. Pose estimation methods provide information on the position and orientation of multiple parts of an animal, with recent methods being able to measure pose information for groups of animals (Chen et al., 2020; Lauer et al., 2021; Pereira et al., 2022; Walter and Couzin, 2021). Some recent methods now even allow for pose estimation to be run in real experimental time (Kane et al., 2020; Lopes et al., 2015; Pereira et al., 2022; Schweihoff et al., 2021).

Methods for pose estimation emerged in computer vision research in the late 1970s (Marr et al., 1978; Nevatia and Binford, 1973). The methods became widely available for the analysis of pose in human behavior following improvements in computer vision (Behnke, 2003), deep learning (Szegedy et al., 2013), and computing using Graphical Processing Units (GPU) (Oh and Jung, 2004). However, these methods were often not robust or required a lot of training data, which were at the time not easily available for animal studies. As a result, a number of open-source tools emerged for pose estimation in animals (e.g., DeepLabCut: Mathis et al., 2018, LEAP: Pereira et al., 2019; DeepPoseKit: Graving et al., 2019). These tools are especially notable in that they were developed to address specific scientific questions by researchers and are not available from commercial sources. They are an outstanding example of the "open source creative process" (White et al., 2019).

One of these methods, DeepLabCut, has been shown to outperform the commercial software package EthoVision XT14 and a hardware-based measurement system from TSE Systems, based on infrared beam breaks (Sturman et al., 2020). When tested across a set of common behavioral assays used in neuroscience, (open field test, elevated plus maze, forced swim test), data from the pose estimation method was evaluated using a neural network classifier and performed as well as classifications by human experts, required data from fewer animals to detect differences due to experimental treatments, and in some cases (head dips in an elevated plus maze) detected effects of treatment (a drug) that was not detected by EthoVision.

In the case of reward seeking behavior, human annotation of videos could resolve the animal's position and when and for how long specific behaviors occurred. These measurements could be made by annotating frames in the video recordings, using tools such as the VIA annotator (Dutta and Zisserman, 2019), and commercial (e.g., EthoVision) or open-source (e.g., ezTrack) methods for whole-animal tracking. These measurements would not be able to account for coordinated movements of multiple body parts or for the dynamics of transitions between different behaviors that together comprise reward seeking behavior. These measurements are easily made using methods for pose estimation. These methods learn to track multiple body parts (for a rodent, the tip of snout, the ears, the base of the tail) and the positions of these body parts can be compared for different kinds of trials (small or large reward) using standard statistical models or machine learning methods. These analyses, together, allow for movements to be categorized (e.g., direct and indirect approach towards a reward port) and for transitions between different types of movements to be quantified (e.g., from turning to walking). It would even be possible to detect unique states associated with deliberation (e.g., head scanning between available choice options). All these measures could then be compared as a function of an experimental manipulation (drug or lesion) or used to assist in the analysis of simultaneously collected electrophysiological or imaging data. None of these measures are possible using conventional methods for annotating video frames or tracking overall the overall position of the animal in a behavioral arena.

Pose estimation methods have been crucial for several recent publications on topics as diverse as tracking fluid consumption to understand the neural coding of reward prediction errors (Ottenheimer et al., 2020), accounting for the effects of wind on the behavior of *Drosophila* (Okubo et al., 2020), understanding the contributions of tactile afferents and nociceptors to the perception of touch in freely moving mice (Schorscher-Petcu et al., 2021), understanding interactions between tactile processing by the rodent whisker system and its ability to guide locomotion (Warren et al., 2021), and measuring the relationship between eye movements and neural activity in freely behaving rodents (Keshavarzi et al., 2022). While a number of studies are emerging that take advantage of methods for pose estimation, there is still not enough widespread adoption of the methods across the research community, perhaps in part due to the technical nature of collecting high quality video recordings as well as setting up and using methods for pose estimation. These methods depend on access to computing systems with GPUs and the ability to set up and use the required computer software, which is usually available as computer code written in Python or MATLAB. A researcher who wants to get started with these approaches will therefore face a number of questions about how to set up video methods in a laboratory setting. New users may also need to learn some of the jargon associated with video analysis methods, and some of these terms are defined in Table 1. *The primary goals of this document are two-fold: to provide information for researchers interested in setting methods for video analysis in a research lab and to propose best practices for the use and development of video analysis methods.*

**A basic setup for video recordings in animal experiments**

In a typical setup for video recording, cameras are placed above, and in some cases to the side or below, the behavioral arena (Figure 1). The cameras send data to a computer and can be integrated with inputs from behavioral devices using custom-written programs using popular libraries such as OpenCV (Bradski, 2000), open-source data collection systems such as Bonsai (Lopes et al., 2015), or software included with many common commercial video capture boards (loopbio Motif). Video files can then be analyzed using a variety of open-source tools.

A common approach is to use methods for pose estimation, which track the position and orientation of the animal. This is done by denoting a set of "keypoints" or "landmarks" (body parts) in terms of pixel locations on frames in the video recordings. Packages for pose estimation provide graphical user interfaces for defining keypoints and the keypoints are then analyzed with video analysis methods. In the example shown in Figure 1, keypoints are the colored dots on the tip of the snout, the ears, forelimbs and paw, midpoint of back, hindlimbs and paws, and base, middle, and end of tail. Once body parts have been defined, computer algorithms are used to track the skeleton formed by the points and to track the position and orientation of the skeleton over frames in the video file. Many open-source tools use machine learning methods for these intensive computational processes, which require GPUs to run in reasonable time. To run these analyses, many labs have either dedicated computers, institutional computing clusters, or cloud computing services such as Google Colab. The outputs of pose estimation can be analyzed to account for movement variability associated with different behaviors, to relate position and orientation to simultaneously collected brain activity (electrophysiology, optical imaging), or with algorithms that can describe and predict states and dynamical transitions of behaviors.

**Data Acquisition:** The first step in setting up for video recording is to purchase a camera with an appropriate lens. Researchers should determine if they need precisely timed video frames, e.g., for integration with electrical or optical recordings. Inexpensive USB webcams with frame rates of at least 30fps are suitable for many neuroscience experiments. However, it is important to make sure that each camera is connected to a dedicated USB channel in the computer used for video recording. Webcam cameras can be a challenge to integrate with systems used for behavioral control and electrophysiology or imaging because they lack a means of precisely synchronizing video frames to other related data. As such, the timing of specific behaviors must be based on the animal's location or an observable event in the video field (e.g., onset of an LED indicating reward availability).

For more precise recordings, specialized cameras used in computer vision applications are needed (e.g., FLIR, Basler). Power and combined data over Ethernet (GigE PoE) is commonly used as it combines long cable length headroom with joint DC power delivery. Alternatively, USB3 cameras can be used, but have a maximum data cable length of 5 m, although active extender cables are available. Most machine vision cameras (GigE PoE or USB3) have general purpose input output (GPIO) capabilities that allow for time synchronization of multiple cameras with other laboratory equipment (e.g., electrical or optical recording system). A single camera running at high resolution or frame rate can quickly saturate a standard 1Gbit Ethernet link. Therefore, it is important to consider the computer used to collect video data, ensuring that it has a fast processor with multiple cores and perhaps also a GPU, which can aid in handling video compression during data collection and can be used for offline analysis using pose estimation methods.

After choosing a camera, researchers must determine how to save and archive data from their recordings. By default, recorded videos from cameras may be in several formats, such as MP4 (MPEG-4 AVC/H. 264 encoding), MOV (MPEG-4 encoding), and AVI (DivX codec, higher quality but larger file size). These formats are generally universal and can be read by a variety of tools. Generally, video data files tend to be large (1 hour of RGB video at 30Hz with resolution 1000×1000 can be 2~20GB depending on compression) so data storage solutions for large-scale experiments are crucial. File compression should be evaluated before a system is deployed, as the computer used for video recordings must have sufficient memory (RAM) to remain stable over long recording sessions. In addition to considerations of file formats and codecs, it is important to plan for data storage. Many labs maintain internal lab servers for their video data. Cloud storage is another option to enable sharing. For sharing data publicly, there are a variety of hosting services available, such as the Open Science Foundation, Figshare, and Dryad (see section on Best Practices below for further comments on data archives and sharing).

Once cameras and lenses are acquired and data formats and storage resolved, the next question is where to position the cameras relative to the experimental preparation. Occlusions due to obstacles, cables, or conspecifics, will have effects on the usability of some video analysis methods. A bottom-up view (from below the animal) works best in an open-field, while a top-down approach can be useful for studies in operant chambers and home cages. Bottom-up views capture behavioral information from the position of the animal's feet (Hsu and Yttri, 2021; Luxem et al., 2022). When multiple cameras are used, to reduce the effect of occlusion for downstream video analysis, cameras should be positioned such that at least one camera can visualize each keypoint at all times.

It is also necessary to think about lighting for the experimental setup. If all or some of the study is to be performed while house lights are off, then infrared (IR) lighting and IR-compatible cameras may be needed. One should consider if diffuse lighting will work or if modifications to eliminate reflective surfaces (especially metals) are necessary. These can lead to artifacts in video recordings from devices like IR LEDs and other sources of illumination and complicate the training and interpretations of measures obtained with analyses such as pose estimation. For example, it is possible to reduce reflections from surfaces and objects that are in direct line with IR LEDs. For top-down recordings, cage floors can be made from colored materials to provide contrast such as Delrin or pre-anodized aluminum (an option for long term use) and the metal pans typically used below operant chambers to collect animal waste can be painted with flat black paint. Addressing these issues before beginning an experiment can greatly improve the quality of video recordings.

Finally, for some applications, it is necessary to invest time in calibrating the video system. Calibration is often overlooked and not easily accessible in many current software packages. The intrinsic parameters of a camera include the focal length of the lens and if the lens has obvious distortions (i.e., fisheye lens). Extrinsic parameters also affect the quality of video recordings and are largely due to the camera's position in the scene. It is fairly easy to calibrate a single camera using a checkerboard or ArUco board. To do so, one sweeps a precalibrated board manually around the field of view of a camera and uses the extracted images to estimate the camera's intrinsic parameters (focal length and distortions). This approach can scale easily to cameras with overlapping fields of view but becomes difficult if larger camera networks do not share extrinsic parameters or need to be repeatedly recalibrated (e.g., if one of the cameras is moved between experiments). If the environment has enough structure in it, structure from motion can estimate the intrinsic and extrinsic parameters by treating the multiple cameras as an exhaustive sweep of the environment. This process can be fully scripted and automatically performed on a daily basis leading to substantially increased reliability and precision in multi-camera system performance. Several references on these topics include Bala et al., 2020, Rameau et al., 2022, Schönberger et al., 2016, and Schonberger and Frahm, 2016.

**Hardware and Software for Data Analysis:** Once video recordings are acquired, the researcher may proceed to setting up their computing environment for pose estimation and tracking. Modern markerless motion capture software tools like DeepLabCut (Mathis et al., 2018) and SLEAP (Pereira et al., 2022) rely on deep learning to automate this process. The most compute-intensive step of these methods involves a "training" stage in which a deep neural network is optimized to learn to predict poses from user-provided examples. Training is typically accelerated with a GPU, a hardware component traditionally used for computer graphics, but which has been co-opted for deep learning due to its massively parallel processing architecture. Having a GPU can speed up training by 10 to 100-fold, resulting in model training times in as little as minutes with lightweight network architectures (Pereira et al., 2022). For most researchers, the most practical option is to purchase a consumer-grade workstation GPU which can be installed in conventional desktop computers to afford local access to this hardware from the pose tracking software. In this case, any recent NVIDIA GPU with greater than 6 GB of memory will suffice for practical use of pose estimation tools. This type of computer hardware has, in recent years, been significantly impacted by supply chain shortages, driving prices up to >$1,000, which makes this a less accessible option for many labs just starting off in video analysis. For this situation, most tools provide the means for using Google Colab, which provides limited access to GPUs on the cloud. This is an excellent way to set up analysis workflows while getting familiar with deep learning-based video analysis but may not be practical for sustained usage (e.g., processing 100s of videos). Another common scenario is that institutions with a high-performance computing center will typically have GPUs available as a local shared resource. Other than GPUs, most other computer requirements are modest (modern CPU, 8-16 GB of RAM, minimal disk space).

Researchers will need to set up their software environment to be able to install and use pose tracking tools. Most commonly available open-source methods for pose estimation were developed using the Python language. It is highly recommended to make use of "environment managers" such as Anaconda ("conda") which enable the creation of isolated installations of Python for each video analysis method of interest. This allows for the methods to be installed with all its dependencies without affecting other Python libraries on the system. Alternatives include Docker, which allows for running an entire virtual machine in isolation. This is done to facilitate the installation of GPU-related dependencies, which may be technically challenging for novice users.

**2D Pose Estimation and Tracking:** Pose tracking methods (Figure 2, part 1) enable researchers to extract positional information about the body parts of animals from video recordings. Tools for pose tracking (see Table 2) decompose the problem of pose tracking into sub-tasks outlined below. A note on nomenclature: pose estimation is the term typically reserved to mean single-animal *keypoint localization* within a single image; multi-animal pose estimation refers to *keypoint localization* and *part grouping* of multiple animals within a single image; and multi-animal pose tracking refers to combined *keypoint localization*, *part grouping* and *identification* across video frames.

Keypoint localization involves recovering the spatial coordinates of each distinct keypoint. This is normally done by estimating body part confidence maps, i.e., image-based representations that encode the probability of the body part being located at each pixel. Recovering the coordinates of each body part is reduced to the task of finding the pixel with highest probability. A key consideration of this task is that the larger the image, the larger the confidence maps. Computer memory requirements can potentially exceed the capacity of most consumer-grade GPUs. This can be compensated by reducing the resolution of the confidence maps, though this comes at the cost of potentially reduced accuracy. Subpixel refinement methods are typically employed to compensate for this, but ultimately confidence map resolution is one of the most impactful choices for achieving reliable keypoint localization.

For single-animal videos, there will be at most *one* instance of each keypoint type present in the image, so keypoint localization is the only step strictly required. For multi-animal videos, however, there may be multiple instances of each keypoint type, e.g., multiple "heads". Part grouping refers to the task of determining the set of keypoint detections that belong to the same animal within an image. This is often approached in either a bottom-up or top-down fashion. In bottom-up models, all parts are detected, the association between them estimated (e.g., by using part affinity fields: Cao et al., 2017), and then grouped. In top-down models, the animals are detected, cropped out of the image, and then keypoints are located in the same fashion as in the single-animal case. These approaches have specific trade-offs. Analyses of bottom-up recordings tend to be more memory-intensive but also more robust to transient occlusions and work well with animals with relatively large bodies (e.g., rodents). By contrast, top-down recordings tend to be analyzed in less time since only subsets of the image are processed. Top-down views work best with smaller body types that have fewer complex occlusions (e.g., flies). A notable consideration is that all single-animal pose estimation models can be used in the multi-animal setting if the animals can be detected and cropped as a preprocessing step (Graving et al., 2019; Pereira et al., 2019). While both methods will work on most types of data, drastic improvements in performance and accuracy can be obtained by selecting the appropriate one – most pose estimation tools allow users to select between each approach type.

Once animals are detected and their keypoints located within a frame, the remaining task in multi-animal pose tracking is identification: repeatedly detecting the same animal across frame sequences. This can be approached as a multi-object tracking (MOT) problem, where animals are matched across frames based on a model or assumption about motion; or a re-identification (ReID) problem, where distinctive appearance features are used to unambiguously identify an animal. Both MOT and ReID (and hybrids) are available as standalone functionality in open-source tools, as well as part of multi-animal pose tracking packages. While MOT-based approaches can function on videos of animals with nearly indistinguishable appearances, they are prone to the error propagation issue inherent in methods with temporal dependencies: switching an animal's identity even once will mean it is wrong for all subsequent frames. This presents a potentially intractable problem for long-term continuous recordings which may be impossible to manually proofread. ReID-like methods circumvent this problem by detecting distinguishing visual features, though this may not be compatible with existing datasets or all experimental paradigms.

The single most significant experimental consideration that will affect the identification problem is whether animals can be visually distinguished. A common experimental manipulation aimed at ameliorating this issue is to introduce visual markers to aid in unique identification of animals. This includes techniques such as grouping animals with different fur colors, painting them with non-toxic dyes (Ohayon et al., 2013), or attaching barcode labels to a highly visible area of their body (Crall et al., 2015). Though an essential part of the pose tracking workflow, identification remains a challenging problem in computer vision and its difficulty should not be underestimated when designing studies involving large numbers of interacting animals. We refer interested readers to previous reviews on multi-animal tracking (Panadeiro et al., 2021; Pereira et al., 2020) for more comprehensive overviews of these topics.

Tools that are based on deep learning work by training deep neural networks (models) to reproduce human annotations of behavior. Methods that strictly depend on learning from human examples are referred to as fully supervised. In the case of animal pose tracking, these supervisory examples (labels) are provided in the form of images and the coordinates of the keypoints of each animal that can be found in them. Most pose tracking software tools fall within this designation and provide graphical interfaces to facilitate labeling. The usability of these interfaces is a crucial consideration as most of the time spent in setting up a pose tracking system will be devoted to manual labeling. The more examples and the greater their diversity, the better that pose tracking models will perform. Previous work has shown that hundreds to thousands of labeled examples may be required to achieve satisfactory results, with a single example taking as much as two minutes to manually label (Mathis et al., 2018; Pereira et al., 2022, 2019). To mitigate this, we strongly recommend adopting a human-in-the-loop labeling workflow. This is a practice in which the user trains a model with few labels, generates (potentially noisy) predictions, and imports those predictions into the labeling interface for manual refinement before retraining the model. This can drastically reduce the amount of time taken to generate thousands of labeled images necessary for reliable pose estimation models.

The rule of thumb is that "if you can see it, you can track it", but this aphorism strongly depends on the examples provided to train the model. Important factors to consider in the labeling stage include labeling consistency and sample diversity. Labeling consistency involves minimizing the variability of keypoint placement within and across annotators which helps to ensure that models can learn generalizable rules for keypoint localization. This can be accomplished by formalizing a protocol for labeling, especially for ambiguous cases such as when an animal's body part is occluded. For example, one convention may be to consistently place a "paw" keypoint at the apex of the visible portion of the body rather than guessing where it may be located beneath an occluding object. Similarly, the principle of consistency should inform *which* body parts are selected as tracked keypoints. Body parts that are not easily located by the human eye will suffer from labeling inconsistency which may cause inferior overall performance as models struggle to find reliable solutions to detecting them. Sample diversity, on the other hand, refers to the notion that not all labeled examples have equal value when training neural networks. For example, labeling 1000 consecutive frames will ensure that the model is able to track data that looks similar to that segment of time, but will have limited capacity to generalize to data collected in a different session. As a best practice, labels should be sampled from the widest possible set of experimental conditions, time points, and imaging conditions that will be expected to be present in the final dataset.

Improving the capability of models to generalize to new data with fewer (or zero) labels is a currently active area of research. Techniques such as transfer learning and self-supervised learning aim to reduce the labeling burden by training models on related datasets or tasks. For example, B-KinD (Sun et al., 2021b) is able to discover semantically meaningful keypoints in behavioral videos using self-supervision without requiring human annotations. These approaches work by training models to solve similar problems and/or on similar data than those used for pose estimation, with the intuition that some of that knowledge can be reused and thereby will require fewer (or no) labeled examples before achieving the same performance as fully supervised equivalents. Future work in this domain is on track to produce reusable models for commonly encountered experimental species and conditions. We highly encourage practitioners to adopt open data and model sharing to facilitate these efforts where possible.

**3D Pose Estimation:** Several methods have emerged in recent years for 3D tracking based on pose data. For some applications, it is of interest to track animals in complete 3D space. This affords a more detailed representation of the kinematics by resolving ambiguities inherent in 2D projections – an especially desirable property when studying behaviors that involve significant out-of-plane movement, such as in large arenas or non-terrestrial behaviors.

It is important to note that 3D motion capture comes at a significant increase in technical complexity. As discussed above (see 'Data Acquisition'), camera synchronization and calibration are paramount for applications using 3D tracking as the result of this step will inform downstream algorithms as to the relative spatial configuration of the individual cameras. This step may be sensitive to small camera movements that occur during normal operation of behavioral monitoring systems, potentially requiring frequent recalibration. The number and positioning of cameras are also major determinants of 3D motion capture performance, both of which may depend on the specific behavior of interest, arena size and bandwidth and computing capabilities on the acquisition computer. In some cases, it may be easiest to use mirrors instead of multiple cameras to allow for recording behavior from multiple perspectives.

Given a calibrated camera system, several approaches have emerged that can enable 3D pose estimation in animals. The simplest approaches rely on using 2D poses detected in each camera view, such as those produced by SLEAP or DeepLabCut as described above, and then *triangulating* them into 3D. 2D poses can be detected by training 2D pose models on each camera view independently, or by training a single model on all views, with varying results depending on how different the perspectives are. Once 2D poses can be obtained, methods such as Anipose (Karashchuk et al., 2021), OpenMonkeyStudio (Bala et al., 2020), and DeepFly3D (Günel et al., 2019) are able to leverage camera calibration information to project poses into 3D for triangulation. This involves optimizing for the best 3D location of each keypoint that still maps back to the detected 2D location in each view. This can be further refined with temporal or spatial constraints, such as known limb lengths. Using this approach, more cameras will usually result in better triangulation, but will suffer (potentially catastrophically) when the initial 2D poses are incorrect. Since many viewpoints will have inherent ambiguities when not all body parts are visible, the 2D pose estimation error issue can be a major impediment to implementing 3D pose systems using the triangulation-based approach.

Alternative approaches attempt to circumvent triangulation entirely. LiftPose3D (Gosztolai et al., 2021) describes a method for predicting 3D poses from single 2D poses, a process known as *lifting*. While this eliminates the need for multiple cameras, it requires a dataset of known 3D poses from which the 2D-3D correspondences can be obtained. This requirement depends on the multi-camera system being similar to the target 2D systems. DANNCE (Dunn et al., 2021), on the other hand, achieves full 3D pose estimation by extending the standard 2D confidence map regression approach to 3D using volumetric convolutions. In their approach, images from each camera view are projected onto a common volume based on the calibration, before being fed into a 3D convolutional neural network that outputs a single volumetric part confidence map. This approach has the major advantage that it is not susceptible to 2D pose estimation errors since it solves for the 3D pose in a single step while also being able to reason about information present in distinct views. The trade-offs with this approach are that it requires significantly more computational power due to the 3D convolutions, as well as requiring 2D ground truth annotations on multiple views for a given frame.

Overall, a practitioner should be mindful of the caveats with implementing 3D pose estimation and is recommended to consider whether the advantages are truly necessary given the added complexity. We note that at the time of writing, none of the above methods can natively support the multi-animal case in 3D, other than by treating them as individual animals after preprocessing with a 2D multi-animal method for pose estimation. This limitation is due to issues with part grouping and identification as outlined above and would seem to be a future area of growth for animal pose estimation.

**Behavior Quantification:** After using pose estimation to quantify the movements of animal body parts, there are a number of analyses that can be used to understand how movements differ by experimental conditions (Figure 2, part 2,3,4). A simple option is to use statistical methods such as ANOVA to assess effects on discrete experimental variables such as the time spent in a given location or the velocity of movement between locations. These measures can also be performed with data from simpler tracking methods, such as the commercially available EthoVision, TopScan, and ANY-maze programs. The primary benefits of the open source pose estimation methods described in this paper over these commercially available programs are the richness of the data obtained from pose estimation (see Figure 1) and the flexibility and customization of behavioral features are tracked (see Figure 2).

If researchers want to go beyond kinematic readouts and investigate the behavior an animal is executing in more detail, then methods for segmenting behavior from the pose tracking data can be used. Behavioral segmentation methods are available to discern discrete episodes of individual events and/or map video or trajectory data to continuous lower-dimensional behavioral representations. Discrete episodes have a defined start and end in which the animal is performing a particular behavior, while continuous representations represent behavior more smoothly over time. For discrete episodes, depending on the experimental conditions, these episodes can last from milliseconds up to minutes or longer. Segmentation can be done per animal e.g., detecting locomotion, or globally per frame, which is especially of interest for social behavior applications. In a global setting researcher might be interested in finding behavioral episodes that are directed between animals such as attacking or mounting behaviors.

If one wants to understand sequences of behaviors, there are many methods available to embed pose data into lower dimensional representations. Such structures can be discovered through unsupervised methods. Some methods provide generic embeddings and do not explicitly model the dynamics of the behaving animal. Two examples of this approach are B-SOiD (Hsu and Yttri, 2021), which analyses pose data with unsupervised machine learning, and MotionMapper (Berman et al., 2014), a method that does not use pose estimation methods. These models embed data points based on feature dynamics (e.g., distance, speed) into a lower dimensional space. Within this space it is possible to apply clustering algorithms for the segmentation of behavioral episodes. Generally, dense regions in this space (regions with many data points grouped together) are considered to be conserved behaviors. Other methods are aimed at explicitly capturing structure from the dynamics (Batty et al., 2019; Bregler, 1997; Costa et al., 2019; Luxem et al., 2022; Shi et al., 2021; Sun et al., 2021a). These models learn a continuous embedding that can be used to identify lower dimensional trajectory dynamics that can be correlated to neuronal activity and segmented in relation to significant behavioral events.

Behavioral segmentation and other methods for quantification require a similar computing environment to that used for pose estimation. The input to those methods is generally the output of a pose estimation method (i.e., keypoint coordinates) or time series from a dimensionality reduction method such as principal component analysis that accounts for the keypoints or the raw video. It is crucial that pose estimation is accurate as the segmentation capabilities of the subsequent methods is bounded by pose tracking quality. Highly noisy key points will drown out biological signals and make the segmentation results hard to interpret, especially for unsupervised methods. Furthermore, identity switches between virtual markers can be catastrophic for multi-animal tracking and segmentation. A summary of methods for behavioral segmentation is provided in Table 3.

Before selecting any approach to segment animal behavior, it is important to first define the desired outcome. If the goal is to identify episodes of well-defined behaviors like rearing or walking, then the most straightforward approach is to use a supervised method. Moreover, it is generally a good starting point to use a supervised learning approach and the outputs of these models can be layered on top of unsupervised models to give them immediate interpretability. One tradeoff, however, is the extensive training datasets that are often required to ensure good supervised segmentation. Such methods can be established quite easily using standard machine learning libraries available for the Python, R, and MATLAB, if one has already experience in building these methods. Alternatively, open-source packages such as SimBA (Nilsson et al., 2020) or MARS (Segalin et al., 2021) can be used, and is especially beneficial for those who are relatively new to the topic of machine learning. However, if the researcher wants to understand more about the spatiotemporal structure of the behaving animal, they either need to label many different behaviors within the video or turn to unsupervised methods. Unsupervised methods offer the advantage to identify clusters in the video or keypoint time series and quantify behavior in each frame. Recently, A-SOiD, an active-learning algorithm, iteratively combines these supervised and unsupervised approaches to reduce the amount of training data required and enable the discovery of additional behavior and structure (Schweihoff et al., 2022).

Interpreting the lower dimensional structures in a 2D/3D projection plot can be difficult and it is advised to visualize examples from this projection space. Generative methods like VAME offer the possibility to sample cluster categories from this embedding space to qualitatively check if similar patterns are learned. Another task unsupervised methods are capable of is fingerprinting. Here, the embedding space is used as a signature to discern general changes in phenotypes (Wiltschko et al., 2020). An alternative to using an explicitly supervised or unsupervised approach is to combine these approaches (semi-supervised), as implemented in a package called TREBA (Sun et al., 2021a). TREBA uses generative modeling in addition to incorporating behavioral attributes, such as movement speed, distance traveled, or heuristic labels for behavior (e.g., sniffing, mounting, attacking) into learned behavioral representations. It has been used in a number of different experimental contexts, most notably for understanding social interactions between animals.

Finally, as behavior is highly hierarchically structured, multiple spatiotemporal scales of description may be desired, e.g., to account for bouts of locomotion and transitions running to escaping behavior (Berman, 2018). It is possible to create a network representation and identify "cliques" or "communities" on the resulting graph (Luxem et al., 2022; Markowitz et al., 2018). These descriptions represent human identifiable behavioral categories within highly interconnected sub-second segments of behavior. These representations can provide insights into the connection between different behavioral states and the transitions between states and their biological meaning.

**Best Practices for Experimenters and Developers**

Having described how to set up and use video recording methods and analysis methods for pose estimation, we would like to close by discussing some best practices in the use and development of methods for video analysis, including recommendations for the open sharing of video data and analysis code.

**Best Practices for Experimenters**: For those using video analysis methods in a laboratory setting, there are several key issues that should be followed as best practices. It is most crucial to develop a means of storing files in a manner in which they can be accessed in the lab, through cloud computing resources, and in data archives. These issues are discussed above in the "Acquisition: Hardware & Software" section of this paper. Documentation of hardware is also a key best practice. All methods sections of manuscripts that use methods for video analysis should include details on the camera and lens that were used, the locations of and distances from the cameras relative to the behavioral arena, the acquisition rate and image resolution, environmental lighting (e.g., IR grids placed above the behavioral arena), properties of the arena (size, material, color, etc.).

Beyond within-lab data management and reporting details on hardware used in research manuscripts, more widespread sharing of video data is very much needed and is a core aspect of best practices for experimenters. In accordance with the demands of funders such as the [NIH](#) for data sharing, the open sharing of raw and processed videos and pose tracking data is crucial for research reproducibility and also for training new users on video methods. Several groups have created repositories to address this need ([Computational Behavior](#), [OpenBehavior](#)). With widespread use, these repositories will help new users learn the required methods for data analysis, enable new analyses of existing datasets that could lead to new findings without having to do new experiments, and would enable comparisons of existing and newly developed methods for pose estimation and behavioral quantification. The latter benefit of data sharing could lead to insight into a major open question about methods for animal pose estimation: how choices about the parameters of any method for pose estimation or subsequent analysis impact analysis time, accuracy, and generalizability. Without these resources, it has not been possible to make confident statements about how existing methods compare across a wide range of datasets involving multiple types of research animals and in different experimental contexts. Guidance for how to implement data sharing can be found in several existing efforts of the machine learning community (Gebru et al., 2021; Hutchinson et al., 2021; Stoyanovich and Howe, 2019). A more widespread use of these frameworks for sharing data can improve the transparency and accessibility of research data for video analysis.

**Best Practices for Developers**: We recommend three topics receive more attention by developers of methods for video analysis. First, there is a need for a common file format for storing results from pose estimation. Second, there is a need for methods to compare pose estimation packages and assess the impact of the parameters of each package on performance in terms of accuracy and user time. Third, there is a need for better code documentation and analysis reproducibility. Each of these issues is discussed below. In addition to these topics, we would like to encourage developers to design interfaces to make their tools more accessible to novice users. This will allow the tools to become more widely used and studied, and will further not limit use of the tools to researchers with advanced technical skills such as programming.

First, it is important to point out that there is no common and efficient data format available for tools that enable pose estimation in animal research. Such a format would allow users to compare methods without having to recode their video data. The FAIR data principles (Wilkinson et al., 2016) are particularly apt for developing a common data format for video due to the large heterogeneity of data sources, intermediate analysis outputs, and end goals of the study. These principles call for data to be Findable (available in searchable repositories and with persistent and citable identifiers (DOIs), Accessible (easily retrieved using the Internet), Interoperable (having a common set of terms to describe video data across datasets), and Reusable (containing information about the experimental conditions and outputs of any analysis or model to allow another group to readily make use of the data). A common file format for saving raw and processed video recordings and data from pose estimation models is needed to address these issues.

Second, there has also been a general lack of direct comparisons of different methods and parameter exploration within a given method on a standard set of videos. The choice of deep learning method and specific hyperparameters can affect the structural biases embedded in video data, thereby affecting the effectiveness of a given method (Sculley et al., 2015). Yet, it seems that many users stick to default parameters available in popular packages. For example, in pose estimation, certain properties of neural network architectures such as its maximum receptive field size can dramatically impact the performance across species owing to the variability in morphological features (Pereira et al., 2022). In addition to the intrinsic properties of particular species (e.g., Hayden et al., 2022), the analysis type will also dictate the importance of particular parameters on the task performance. For example, algorithms that achieve temporal smoothness in pose tracking are crucial for studies of fine motor control (Wu et al., 2020), but perhaps not as essential as preventing identity swaps for studies of social behavior (Pereira et al., 2022; Segalin et al., 2021). Another important issue is that most methods do not report well-calibrated measures of the confidence of model fits or predictions. This is important as it has become clear that machine learning tools tend to be overconfident in their predictions (Abdar et al., 2021). Establishing standardized, interoperable data formats and datasets that include estimates of the fitted models and their predictions will enable comprehensive comparisons of existing and new methods for pose estimation and behavioral quantification.

For evaluating specific methods on lab-specific data, appropriate metrics and baseline methods for the research questions should be chosen. There may be cases where comparable baseline methods may not exist. For example, if a lab develops a new method for quantifying behavior for a specific organism or task on a lab-specific dataset, and there are no existing studies for that task. However, if related methods exist, it would be beneficial to compare performance of the new method against existing methods to study the advantages and disadvantages of the method. For more general claims (e.g., state-of-the-art pose estimator across organisms), evaluations on existing datasets and comparisons with baselines is important (see Table 4), to demonstrate the generality of the method and improvements over existing methods. A consensus on a standard set of data in the community for evaluation and an expansion to include more widely used behavioral tasks and assays would facilitate general model development and comparison. We show existing datasets in the community for method development in Table 4 and encourage the community to continue to open-source data and expand this list of available datasets to accelerate model development.

Third, reproducibility of results is crucial for acceptance of new methods for video analysis within the research community and for research transparency. Guidance for documenting the details of models and algorithms can be obtained from the [Machine Learning Reproducibility Checklist](). It is applicable to any computational model in general. Importantly, the checklist calls for including the range of hyperparameters considered for experiments, mean and variance of results from multiple runs, and an explanation of how samples were allocated for train/validation/test. Further guidance for sharing code is available in this GitHub resource: [Publishing Research Code](). It provides tips on open-sourcing research code, including specifications of code dependencies, training and evaluation code, and including pre-trained models as part of any code repository. Beyond these resources, we note that there is also a broader definition of reproducibility in that experiments should be *robustly reproducible*: experimental results should ideally not vary significantly under minor perturbations. For example, even if there are minor variations to lighting or arena size from the original experiments, the video analysis results should not change significantly. A framework to ensure robust reproducibility is currently an open question, but the existing frameworks should facilitate producing the same results under the same experimental conditions. *Model interpretability* is another important consideration depending on the purpose of the video analysis experiment. Many machine learning models are "black box" models, and not easily interpretable; as such, post-hoc explanations may not always be reliable (Rudin, 2019). One way to generate human-interpretable models is through program synthesis (Balog et al., 2017) and neurosymbolic learning (Sun et al., 2022; Zhan et al., 2021). These methods learn compositions of symbolic primitives, which are closer in form to human-constructed models than neural networks. Interpretable models can facilitate reproducibility and trustworthiness in model predictions for scientific applications. Efforts at deploying these approaches for methods for video analysis and behavioral quantification are very much needed.

**Summary**

We hope that our review of the current state of open-source tools for behavioral video analysis will be helpful to the community. We described how to set up video methods in a lab, provided an overview on currently available methods, and provided guidance for best practices in using and developing the methods. As newer tools emerge and more research groups become proficient at using available methods, there is a clear potential for the tools to help with advancing our understanding of the neural basis of behavior.

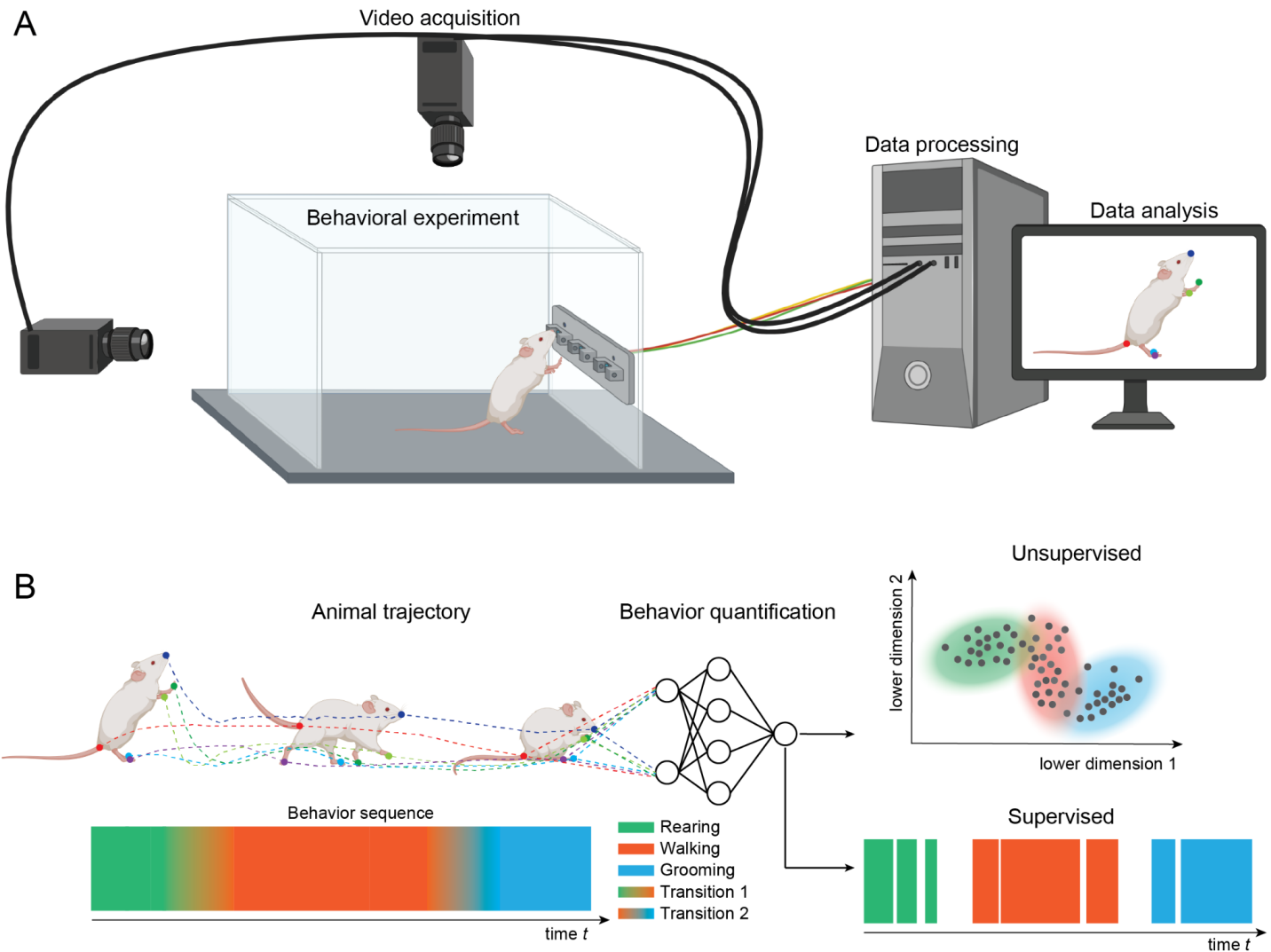

**Figure 1: Setup for video recording.** (A) Cameras are mounted above and to the side of a behavioral arena. The cameras record sequences of images of an animal performing a behavioral task. The recordings are stored on a computer and analyzed with methods for pose estimation and behavior classification. (B) The animal's pose trajectory captures the relevant kinematics of the animal's behavior and is used as input to behavior quantification algorithms. Quantification can be done using either unsupervised (learning to recognize behavioral states) or supervised (learning to classify behaviors based on human annotated labels). In this example, transitions among three example behaviors (Rearing, Walking and Grooming) are depicted on the lower left and classification of video frames into the three main behaviors are depicted on the lower right.

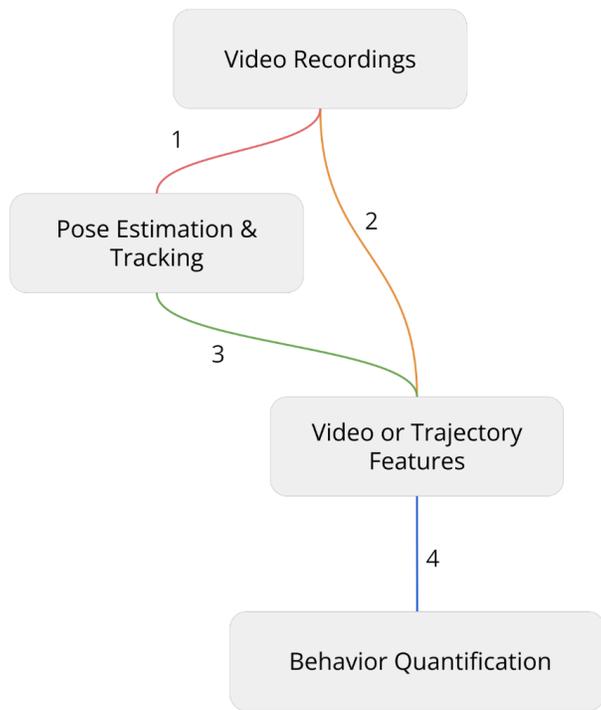
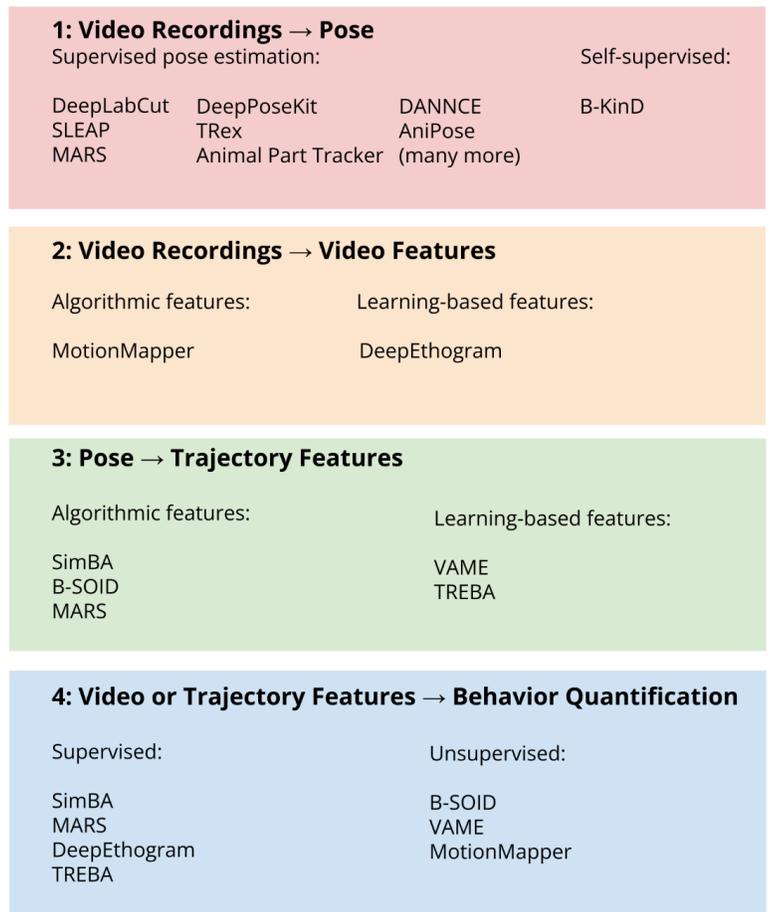

**Figure 2: Pipeline for video analysis.** Video recordings are analyzed with either keypoints from 2D or 3D pose estimation or directly by computing video features. These video or trajectory features are then used by downstream algorithms to relate the keypoints to behavioral constructs such as predicting human-defined behavior labels (supervised learning) or discovering behavior motifs (unsupervised learning). Each part of the analysis steps outlined in the figure is described in more detail below.

**Table 1: Frequently used terms for video analysis**

| | |
|---|---|
| pose | The configuration (position and/or orientation) of an animal, object, or body parts in an image or video recording |
| keypoints/landmarks | Distinct identifiable morphological features (e.g., the tip of the snout or the base of the tail in a rodent) that can be localized in 2D or 3D from images, typically via pose estimation |
| part grouping | A process for assigning keypoints to individual animals |
| multi-object tracking | In multi-animal pose tracking, the task of determining which detected poses belong to which individual animal across time |
| re-identification | A process for identifying all images containing the same individual animal based primarily on their distinct appearance |
| kinematics | Information about the angles and velocities of a set of keypoints |
| supervised learning | Machine learning methods that use experimenter-provided labels (e.g., ground truth poses, or 'running' vs 'grooming') to train a predictive model |
| unsupervised learning | Machine learning methods that only use unlabeled data to find patterns based on its intrinsic structure (e.g., clustering behavioral motifs based on the statistics of their dynamics) |
| transfer learning | Machine learning methods that use models trained on one dataset to analyze other datasets (e.g., models of grooming in mice applied to rats) |
| self-supervised learning | Machine learning methods that use only unlabeled data for training by learning to solve artificially constructed tasks (e.g., comparing two variants of the same image with noise added against other images; predicting the future; or filling in blanks) |
| embedding | A representation of high-dimensional data into lower dimensional representation |
| lifting | A process through which 2D pose data are converted to 3D representations |
| behavioral segmentation | A process for detecting occurrences of behaviors (i.e., starting and ending frames) from video or pose sequences |

**Table 2: Methods for 2D pose estimation**

| | |
|---|---|
| DeepLabCut | DeepLabCut (Mathis et al., 2018) uses a popular architecture for deep learning (He et al., 2016), called *ResNet*. DeepLabCut models are pre-trained on a massive dataset for object recognition called ImageNet (Russakovsky et al., 2015). Through a process called *transfer learning*, the DeepLabCut model learns the position of keypoints using as few as 200 labeled frames. This makes the model very robust and flexible in terms of what body parts (or objects) users want to label as the model provides a strong backbone of image filters within their ResNet architecture. To detect the keypoint position, DeepLabCut replaces the classification layer of the *ResNet* with deconvolutional layers to produce spatial probability densities from which the model learns to assign high probabilities to regions with the user labeled keypoints. DeepLabCut can provide very accurate pose estimations but can require extensive time for training. https://github.com/DeepLabCut/DeepLabCut |
| SLEAP | SLEAP (Pereira et al., 2022) is based on an earlier method called LEAP (Pereira et al., 2019), which performed pose estimation on single animals. SLEAP uses simpler CNN architectures with repeated convolutional and pooling layers. This makes the model more lightweight compared to DLC's ResNet architecture and, hence, the model is faster to train with comparable accuracy. Similar to DeepLabCut, the model uses a stack of upsampling or deconvolutional layers to estimate confidence maps during training and inference. Unlike DLC, SLEAP does not solely rely on transfer learning from general-purpose network models (though this functionality is also provided for flexible experimentation). Instead, it uses customizable neural network architectures that can be tuned to the needs of the dataset. SLEAP can produce highly accurate pose estimates starting at about 100 labeled frames for training combined and is quick to train on a GPU (<1 hour). https://sleap.ai/ |
| DeepPoseKit | DeepPoseKit (Graving et al., 2019) uses a type of CNN architecture, called stacked DenseNet, an efficient variant of the stacked hourglass (Newell et al., 2016), and uses multiple down- and upsampling steps with densely connected hourglass networks to produce confidence maps on the input image. The model uses only about 5% of the amount of parameters used by DeepLabCut, providing speed improvements over DeepLabCut and LEAP. https://github.com/jgraving/DeepPoseKit |
| B-KinD | B-KinD (Sun et al., 2021b) discovers key points without human supervision. B-KinD has the potential to transform how pose estimation is done, as keypoint analysis is one of the most time-consuming aspects of doing pose estimation analysis. However, there are challenges for the approach when occlusions occur in the video recordings, e.g., recordings of animals tethered to brain recording systems. https://github.com/neuroethology/BKinD |

**Table 3: Methods for behavioral segmentation using pose data**

| | |
|---|---|
| SimBA | SimBA (Nilsson et al., 2020) is a supervised learning pipeline for importing pose estimation data and a graphical interface for interacting with a popular machine learning algorithm called Random Forest (Breiman, 2001). SimBA was developed for studies in social behavior and aggression and has been shown to be able to discriminate between attack, pursuit, and threat behaviors in studies using rats and mice.<br>https://github.com/sgoldenlab/simba |
| MARS | MARS (Segalin et al., 2021) is another supervised learning pipeline developed for studies of social interaction behaviors in rodents, such as attacking, mounting, and sniffing, and uses the XGBoost gradient boosting classifier (Chen and Guestrin, 2016).<br>https://github.com/neuroethology/MARS |
| B-SOiD | B-SOiD (Hsu and Yttri, 2021) uses unsupervised methods to learn and discover the spatiotemporal features in pose data of ongoing behaviors, such as grooming and other naturalistic movements in rodents, flies, or humans. B-SOiD uses UMAP embedding (McInnes et al., 2020) to account for dynamic features within video frames that are grouped using an algorithm for cluster analysis, HDBSCAN (McInnes et al., 2017). Clustered spatiotemporal features are then used to train a classifier (Random Forest; Breiman, 2001) to detect behavioral classes in data sets that were not used to train the model and with millisecond precision.<br>https://github.com/YttriLab/B-SOID |
| VAME | VAME (Luxem et al., 2022) uses self-supervised deep learning models to infer the full range of behavioral dynamics based on the animal movements from pose data. The variational autoencoder framework (Kingma and Welling, 2019) is used to learn a generative model. An encoder network learns a representation from the original data space into a latent space. A decoder network learns to decode samples from this space back into the original data space. The encoder and decoder are parameterized with recurrent neural networks. Once trained, the learned latent space is parameterized by a Hidden Markov Model to obtain behavioral motifs.<br>https://github.com/LINCellularNeuroscience/VAME |
| TREBA | TREBA (Sun et al., 2021a) relates measures from pose estimation to other quantitative or qualitative data associated with each frame in a video recording. Similar to VAME, a neural network is trained to learn to predict movement trajectories in an unsupervised manner. TREBA can then incorporate behavioral attributes, such as movement speed, distance traveled, and heuristic labels for behavior (e.g., sniffing, mounting, attacking) into representations of the pose estimation data learned by its neural networks, thereby bringing aspects of supervised learning. This is achieved using a technique called task programming.<br>https://github.com/neuroethology/TREBA |

Table 4: Datasets for model development

| Dataset | Task | Setting | Organism |
|---|---|---|---|
| Human3.6M | 2D/3D Pose Estimation | Videos from 4 camera views with poses from motion capture | Human (single-agent) |
| MS COCO | 2D Pose Estimation | Images from uncontrolled settings with annotated poses | Human (multi-agent) |
| PoseTrack | 2D Pose Estimation & Tracking | Videos from crowded scenes with annotated poses | Human (multi-agent) |
| AP-10K | 2D Pose Estimation | Images of diverse animal species with annotated poses | Diverse species (single & multi-agent) |
| MARS | 2D Pose Estimation | Videos from 2 camera views with annotated poses | Mouse (multi-agent) |
| 3D-ZEF | 2D/3D Pose Estimation & Tracking | Videos from 2 camera views with annotated poses | Zebrafish (multi-agent) |
| OpenMonkeyStudio | 2D/3D Pose Estimation | Images with annotated poses from a 62 camera setup | Monkey (single-agent) |
| PAIR-R24M | 2D/3D Pose Estimation & Tracking | Videos from 12 camera views with poses from motion capture | Rat (multi-agent) |
| 3DPW | 2D/3D Pose Estimation & Tracking | Videos from moving phone camera in challenging outdoor settings | Human (multi-agent) |
| 3DHP | 2D/3D Pose Estimation | Videos from 14 camera views with poses from motion capture | Human (single-agent) |
| Rat 7M | 2D/3D Pose Estimation | Videos from 12 camera views with poses from motion capture | Rat (single-agent) |

| Dataset | Task | Setting | Organism |
|---|---|---|---|
| Kinetics | Video-level Action Classification | Videos from uncontrolled settings that cover 700 human actions | Human (single & agent, may interact with other organisms/objects) |
| NTU-RGBD | Video-level Action Classification (also has 3D poses) | Videos from 80 views and depth with 60 human actions | Human (single & multi-agent) |
| MultiTHUMOS | Frame-level Action Classification | Videos from uncontrolled settings with 65 action classes | Human (single & multi-agent) |
| CRIM13 | Frame-level Behavior Classification | Videos from 2 views, with 13 annotated social behaviors | Mouse (multi-agent) |
| Fly vs. Fly | Frame-level Behavior Classification (also has 2D poses) | Videos & trajectory, with 10 annotated social behaviors | Fly (multi-agent) |
| CalMS21 | Frame-level Behavior Classification (also has 2D poses) | Videos & trajectory, with 10 annotated social behaviors | Mouse (multi-agent) |
| MABe | Frame-level Behavior Classification (also has 2D poses) | Top-down views, 7 annotated keypoints, hundreds of videos | Mouse (multi-agent) |